\documentclass[aps,showpacs,twocolumn,superscriptaddress,pra]{revtex4}

\usepackage{amsmath}
\usepackage{amssymb}
\usepackage{amscd}
\usepackage{wasysym}
\usepackage[ansinew]{inputenc}
\usepackage[T1]{fontenc}
\usepackage{ae,aecompl}

\usepackage[dvips]{graphicx}

\newcommand{\be}{\begin{equation}}
\newcommand{\ee}{\end{equation}}
\newcommand{\bea}{\begin{eqnarray}}
\newcommand{\eea}{\end{eqnarray}}
\newcommand{\ba}{\begin{align}}
\newcommand{\ea}{\end{align}}
\newcommand{\nn}{\nonumber}

\newcommand{\ket}[1]{|#1\rangle}
\newcommand{\bra}[1]{\left\langle#1\right|}

\newcommand{\eye}{\mbox{$\mbox{1}\!\mbox{l}\;$}}

\newcommand{\tr}{{\rm tr}}

\begin{document}
\bibliographystyle{apsrev}

\title{Dissipation induced macroscopic entanglement in an open optical lattice}

\author{G. Kordas}
\affiliation{Institut f\"ur theoretische Physik and Center for Quantum
             Dynamics, Universit\"{a}t Heidelberg, Heidelberg, Germany}
\affiliation{University of Athens, Physics Department, 
		Panepistimiopolis, Ilissia 15771 Athens, Greece}
\author{S. Wimberger}
\affiliation{Institut f\"ur theoretische Physik and Center for Quantum
             Dynamics, Universit\"{a}t Heidelberg, Heidelberg, Germany}
\author{D. Witthaut}
\affiliation{Network Dynamics, Max Planck Institute for Dynamics and Self-Organization,
		D--37077 G\"ottingen, Germany}

\begin{abstract}
We introduce a method for the dissipative preparation of strongly
correlated quantum states of ultracold atoms in an optical lattice
via localized particle loss.
The interplay of dissipation and interactions enables different types
of dynamics. This ushers a new line of experimental methods to maintain 
the coherence of a
Bose-Einstein condensate or to deterministically generate
macroscopically entangled quantum states.
\end{abstract}


\maketitle


\section{Introduction}

Decoherence and dissipation, caused by the irreversible coupling of a 
quantum system to its environment, represent a major obstacle for a 
long-time coherent control of quantum states.
Sophisticated methods have been developed to maintain coherence 
in open quantum systems with applications in quantum control
and quantum information processing \cite{Niel00,Gard04}.
Only recently a new paradigm has been put forward: 
Dissipation can be used as a powerful tool to steer the dynamics
of complex quantum systems if it can be accurately controlled.
It was shown that dissipative processes can be tailored
to prepare arbitrary pure states \cite{Diel08,Diel11} or to enable 
universal quantum computation \cite{Vers09}.
Several methods have been proposed to dissipatively
generate entangled states \cite{Kast11,Krau10}.
However, most proposals rely on rather special dissipation processes,
such that a sophisticated control of the coupling of system and 
environment must be realized. 

In this letter we propose a scheme to create macroscopically 
entangled quantum states of ultracold atoms based on localized 
particle loss. This process can be readily realized in ongoing 
experiments with optical lattices enabling a single-site access 
\cite{Albi05,Gros10,Bakr09,Sher10,Geri08,Wurt09}.
The generated quantum states show remarkable statistical 
properties: The atoms relax to a coherent superposition of
bunches localized at different lattice positions. These states 
generalize the so-called NOON states enabling interferometry 
beyond the standard quantum limit \cite{Boll96,Giov04}. Furthermore, 
they may serve as a distinguished probe of decoherence and the 
emergence of classicality.
As particle loss is an elementary and omnipresent dissipation process,
this method may be generalized to a variety of open quantum systems
well beyond the dynamics of ultracold atoms.

\section{Particle loss in an optical lattice}

The coherent dynamics of bosonic atoms in an optical lattice is described 
by the Bose-Hubbard Hamiltonian \cite{Jaks98}
\be
  \hat H = - J  \sum \nolimits_{j} \left( \hat a_{j+1}^{\dagger} \hat a_j +
                  \hat a_{j}^{\dagger} \hat a_{j+1} \right)
         + \frac{U}{2} \sum \nolimits_j 
           \hat a_{j}^{\dagger}  \hat a_{j}^{\dagger} 
               \hat a_{j}  \hat a_{j},
    \label{eqn-hami-bh}
\ee
where $\hat a_j$ and $\hat a_j^\dagger$ are the bosonic 
annihilation and creation operators in the $j$th well. We set $\hbar = 1$, thus measuring all energies 
in frequency units. Throughout this letter we assume periodic
boundary conditions.

\begin{figure*}[tb]
\centering
\includegraphics[width=4.4cm, angle=0]{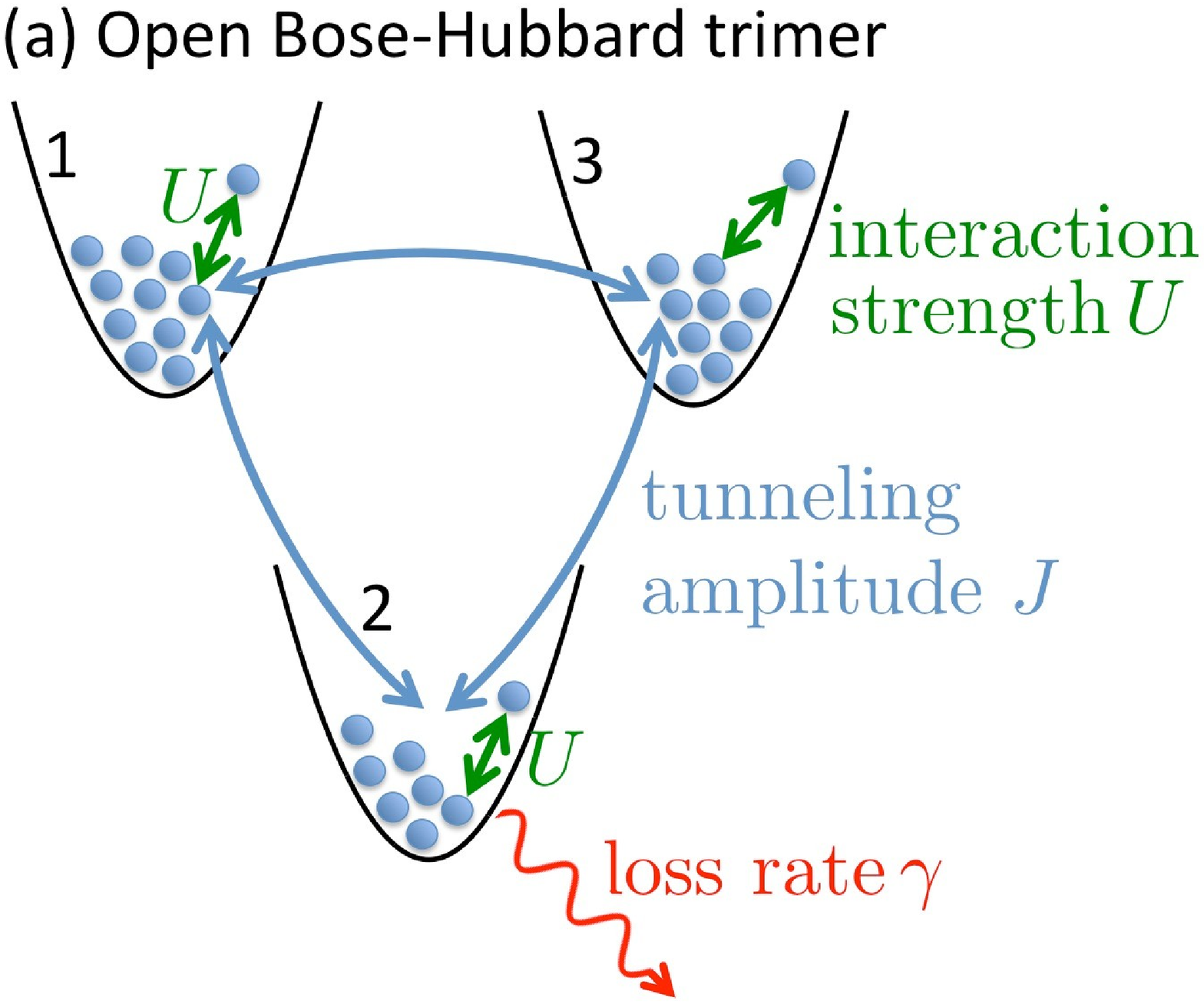}
\hspace{4mm}
\includegraphics[width=11.2cm, angle=0]{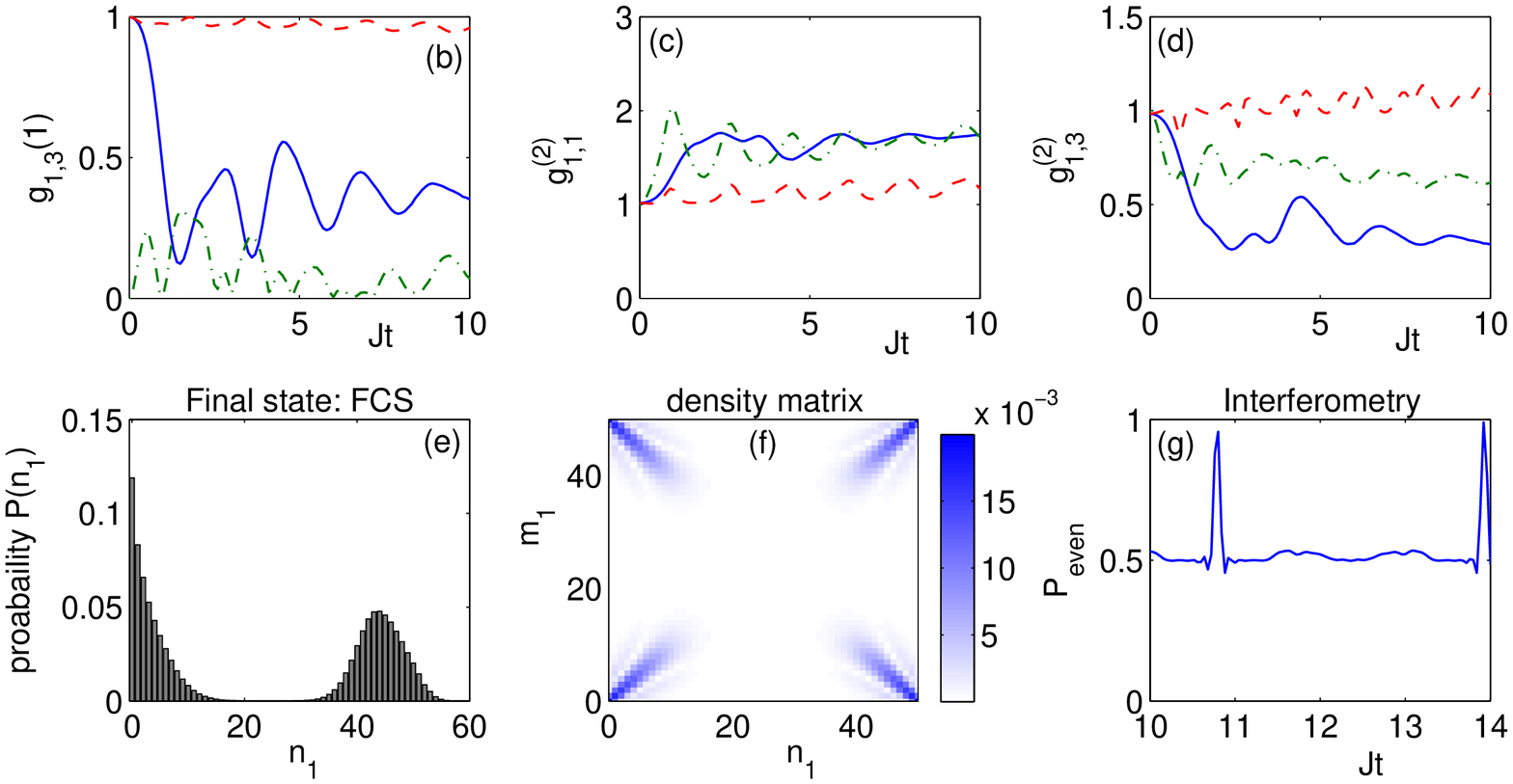}
\caption{\label{fig:trimer}
Dissipative generation of a macroscopically entangled breather state
in an open Bose-Hubbard trimer. 
(a) Schematic drawing of the system.
(b) Evolution of the phase coherence $g_{1,3}^{(1)}(t)$
for the symmetric initial state $\ket{\Psi_+}$ (dashed line),
the anti-symmetric initial state $\ket{\Psi_-}$ (solid line),
and the Fock state $\ket{\Psi_F}$ (dash-dotted line).
(c,d) Evolution of the number fluctuations  $g_{1,1}^{(2)}(t)$ 
and correlations  $g_{1,3}^{(2)}(t)$.
(e,f) Analysis of the final state at $t=10 J^{-1}$ for the 
anti-symmetric initial state $\ket{\Psi_-}$.
(e) The full counting statistics in the first well. 
(f) The density matrix elements $\rho(n_1,0,n-n_1;m_1,0,n-m_1)$ for
$n=50$ reveals the coherence of the breather state.
(g) Matter wave interferometry of the breather state.
Plotted is the probability to detect an even number of atoms at site 1.
Parameters are $U = 0.1 J$, $\gamma_2 = 0.2 J$, 
$\gamma_1 = \gamma_3 = 0$ 
and $N(0) = 60$. 
}
\end{figure*}

We analyze the dynamics of ultracold atoms induced
by particle loss from a single lattice site acting in concurrence with
strong atom-atom interactions. Single site access can be implemented 
optically either by increasing the lattice period \cite{Albi05,Gros10}
or by pushing the resolution of the optical imaging system to the limit
\cite{Bakr09,Sher10}. 
Detection and coherent manipulation of the atoms with off-resonant light have been demonstrated, whereas a controllable particle loss can be implemented with a strong resonant blast laser in a straightforward way.
An even higher resolution can be realized by a focussed electron beam ionizing atoms which are then removed from the lattice \cite{Geri08,Wurt09}. This can be used for the detection of single atoms as well as for an incoherent manipulation of the quantum dynamics in the lattice \cite{Braz09,11leaky1,11leaky2}.

In addition to this tunable source of dissipation, phase noise can limit 
the coherence in the lattice 
\cite{Angl97,Truj09,Pich10,Gerb10}. The dynamics 
of the atoms is then described by a quantum master equation 
\cite{Gard04,11leaky1,11leaky2,Barm11,Kepe12}
\bea
  \dot{\hat \rho} &=& -i [\hat H,\hat \rho] 
  - \frac{\kappa}{2} \sum_{j = 1,2}
    \left( \hat n_j^2 \hat \rho + \hat \rho \hat n_j^2 
          - 2 \hat n_j \hat \rho \hat n_j  \right) \nn \\
   && \quad - \frac{1}{2}  \sum_{j}  \gamma_{j} \left(
     \hat a_j^\dagger \hat a_j \hat \rho + \hat \rho \hat a_j^\dagger \hat a_j  
    - 2 \hat a_j \hat \rho \hat a_j^\dagger \right),
   \label{eqn:master2}
\eea
where $\gamma_j$ is the loss rate at the $j$th site, $\hat n_j = \hat a_j^\dagger \hat a_j$ are the number 
operators and $\kappa$ is the
rate of phase noise. We analyze the effects of decoherence
in Fig.~\ref{fig:ent}, otherwise we set $\kappa=0$. 
Numerical simulations are carried
out using the quantum jump method \cite{Dali92} for small systems and
the truncated Wigner method \cite{Sina02} for large lattices.

\section{Dissipation induced macroscopic entanglement}

We first consider the dynamics of ultracold atoms in a triple-well trap as illustrated
in Fig.~\ref{fig:trimer} (a), where a numerically exact solution is still possible
for reasonable atom numbers. We analyze the quantum state of the atoms
characterized by the correlation functions 
\be
     g^{(1)}_{j,\ell} = \frac{\langle  \hat a_j^\dagger \hat a_\ell \rangle }{
                             \sqrt{ \langle \hat n_j \rangle \langle \hat n_\ell \rangle}}
      \qquad \mbox{and} \qquad                       
      g^{(2)}_{j,\ell} = \frac{ \langle  \hat n_j \hat n_\ell \rangle }{
                             \langle \hat n_j \rangle \langle \hat n_\ell \rangle}.
      \label{eqn:g12}                       
\ee  
The first-order correlation function $g^{(1)}$ measures the phase coherence
between two wells, while $g^{(2)}$ gives the number fluctuations or number 
correlations between different wells.
The evolution of these functions is plotted in Fig.~\ref{fig:trimer} (b-d)
for three different initial states, a pure Bose-Einstein condensate (BEC) with
(anti-)symmetric wave function $\ket{\Psi_\pm} \sim 
(\hat a_1^\dagger \pm \hat a_3^\dagger)^N \ket{0,0,0}$
and a Fock state $\ket{\Psi_F} = \ket{N/2,0,N/2}$, respectively. 
In all cases we assume strong atom-atom interactions $U = 0.1 J$, such that $UN>J$.  
For the symmetric initial state $\ket{\Psi_+}$, all correlations 
remain close to the initial values indicating that the BEC 
remains approximately pure for all times. 
More precisely, the quantum state is approximately given by a 
superposition of pure product states $\ket{\Psi_n} = \ket{\psi_1}^{\otimes n}$ with different atom numbers $n$, 
where all atoms occupy the same single-particle 
state $\ket{\psi_1}$.
The atoms decay from the lattice in an uncorrelated way, which is well
described by mean-field theory \cite{11leaky1,11leaky2}.
Another more subtle effect is that localized particle loss
can maintain or even restore the purity of a condensate 
as non-condensed atoms are rapidly removed from the lattice
\cite{08mfdecay,08stores}. 

In contrast, the anti-symmetric initial state $\ket{\Psi_-}$ is 
dynamically unstable such that the condensate
is rapidly destroyed and phase coherence is lost. 
The nature of the emerging quantum state is revealed by the 
$g^{(2)}$-function:
Number fluctuations strongly increase while the correlations decrease.
This shows that the atoms start to bunch at one lattice site while the
other sites are essentially empty. A similar dynamics is found for the Fock
state $\ket{\Psi_F}$, however the number 
anti-correlations are less
pronounced and the equilibration to the final state takes a longer
time. The emerging state is called a breather state in the following
in analogy to localized modes in nonlinear lattices \cite{Camp04,Ng09}. 

The full counting statistics of the atoms shown in 
Fig.~\ref{fig:trimer} (e) clearly reveals that the atoms 
relax either to site $1$ or to site $3$, leaving site $2$
essentially empty. Most interestingly, these two 
contributions are phase coherent, which is
confirmed by an analysis of the density matrix 
of the atoms. Figure \ref{fig:trimer} (f)
shows the matrix elements $\rho(n_1,0,n-n_1;m_1,0,n-m_1)$ for
$n=50$, the most probable value of the atom number
at $t=10J^{-1}$. 
Full coherence is observed between the contributions with small 
and large atom number at site $1$, i.e. $n_1 \apprge 0$ and 
$n_1 \apprle 50$.

The breather states generated by this protocol generalize the so-called
NOON states $\ket{n,0,0} + e^{i \vartheta} \ket{0,0,n}$ which enable precision
interferometry beyond the standard quantum limit \cite{Giov04}. 
Breather states can be written as a superposition of states of the form 
\be
     \ket{n_1,n_2,n-n_1-n_3} + e^{i \vartheta} \ket{n-n_1-n_3,n_2,n_1}. 
\ee
The number of atoms $n$ varies, but the coherence of wells 
$1$ and $3$ is guaranteed, which
is sufficient for precision interferometry. 
We consider an interferometric measurement, where the modes (lattice 
sites) $1$ and $3$ are mixed as given by the time evolution operator
$\hat{U} = \exp(-i\hat{H}_{\rm mix} t)$ with 
$\hat{H}_{\rm mix} = i J(\hat a_1^\dagger \hat a_3 - \hat a_3^\dagger \hat a_1)$,
assuming that  interactions and loss are switched off.
In analogy to the parity observable \cite{Boll96},
we record the probability $P_{\rm even}(t)$
to detect an even number of atoms 
 in lattice site $1$,
which is shown as a function of time in Fig.~\ref{fig:trimer} (g).  
We find that $P_{\rm even}$ approaches unity periodically,
which proves that the breather states are
fully phase coherent and thus enable quantum interferometry
as ordinary NOON states. Breather states are readily 
generated for large particle numbers, which is 
notoriously difficult using other methods (see, e.g., \cite{Afek10}).

\section{Entanglement and decoherence}

The atoms in a breather or NOON state are strongly entangled:
If some atoms are measured at one site, then the remaining 
atoms will be projected to the same site with overwhelming probability.
To unambiguously detect this form of multi-partite entanglement,
we analyze the variance of the population imbalance 
$\Delta(\hat n_3 - \hat n_1)^2$, which scales as $\sim N^2$
for a breather state, while it is bounded by $N$ for a pure product state,
$N$ being the total atom number.
Given a pure state decomposition of the quantum state 
$\hat \rho = L^{-1} \sum_{a=1}^L \ket{\psi_a}\bra{\psi_a}$, 
we introduce the entanglement parameter
\bea
       \label{eqn:ent_para}    
    E_{j,k} &:=& \langle (\hat n_j - \hat n_k)^2 \rangle
     - \langle \hat n_j - \hat n_k \rangle^2 
     - \langle \hat n_j + \hat n_k \rangle  \\
    &&  - \frac{1}{2L^2} \sum_{a,b} 
     \left[ \langle \hat n_j - \hat n_k \rangle_a 
         - \langle \hat n_j - \hat n_k \rangle_b \right]^2, \nn
\eea
for the wells $(j,k)$,
where $\langle \cdot \rangle_{a,b}$ denotes the expectation value in 
the pure state $\ket{\psi_{a,b}}$. Such a pure state decomposition is
automatically provided by a quantum jump simulation \cite{Dali92}. 
The last term in the parameter $E$ corrects 
for the possibility of an incoherent superposition of states localized at
sites $1$ and $3$. For a separable quantum state one can show that 
$E_{j,k}<0$ such that a value $E_{j,k}>0$ unambiguously proves
entanglement  of the atoms. A proof is given in the appendix.

\begin{figure}[tb]
\centering
\includegraphics[width=8cm, angle=0]{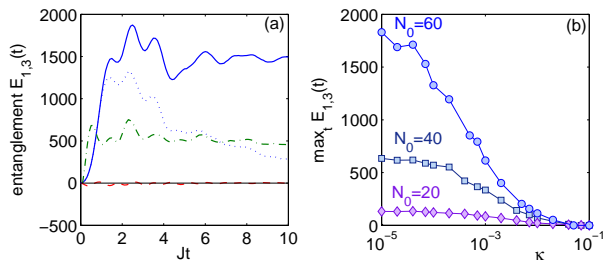}
\caption{\label{fig:ent}
Entanglement and decoherence of a breather state in a triple-well trap.
(a) Evolution of the entanglement parameter (\ref{eqn:ent_para})
for $\kappa=0$ and three different initial states, 
$\ket{\Psi_+}$ (dashed line), $\ket{\Psi_-}$ (solid line),
and $\ket{\Psi_F}$ (dash-dotted line),
and for  $\ket{\Psi_-}$ in the presence
of modest phase noise, $\kappa = 10^{-4} J$ (dotted line)
and $N(0) = 60$.
(b) Maximum of the entanglement parameter ${\rm max}_t E_{1,3}(t)$
as a function of the noise rate $\kappa$ for the anti-symmetric
initial state $\ket{\Psi_-}$ and different particle numbers.
The remaining parameters are the same as in Fig.~\ref{fig:trimer}.
}
\end{figure}

As shown in Fig.~\ref{fig:ent} (a),
$E_{1,3}(t)$ rapidly relaxes to a large non-zero value
for a Fock or an anti-symmetric  initial state, which is 
maintained during the full duration of the simulation. This proves the 
deterministic generation of a meta-stable macroscopically entangled 
quantum state by localized particle dissipation. 
Furthermore, entangled breather states provide a sensitive probe 
for environmentally induced decoherence. Figure \ref{fig:ent} (b) 
shows the maximum value of $E_{1,3}(t)$ realized in the presence of 
phase noise. Entanglement decreases with the noise rate $\kappa$,
in which breather states with large particle numbers are most sensitive.
However, entanglement persists up to relatively large values of 
$\kappa \approx 10^{-2}J$ in all cases.

\begin{figure}[tb]
\centering
\includegraphics[width=8cm, angle=0]{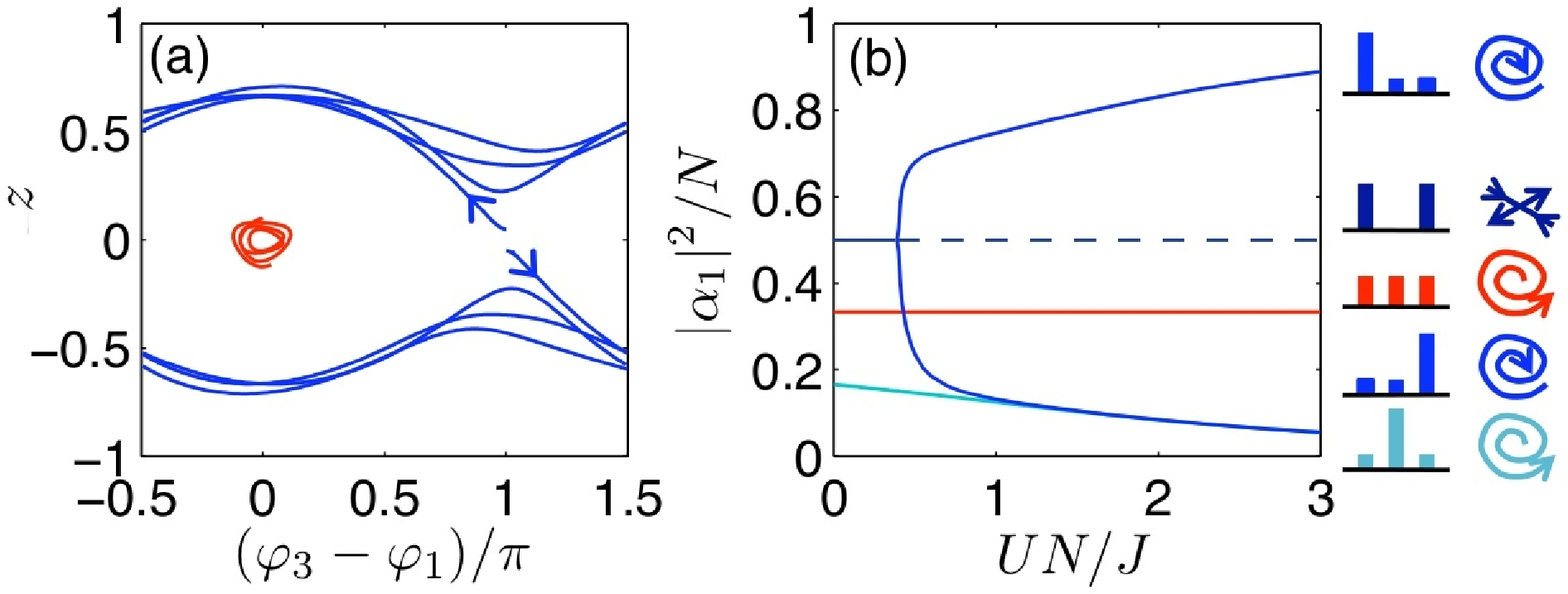}
\includegraphics[width=4cm, angle=0]{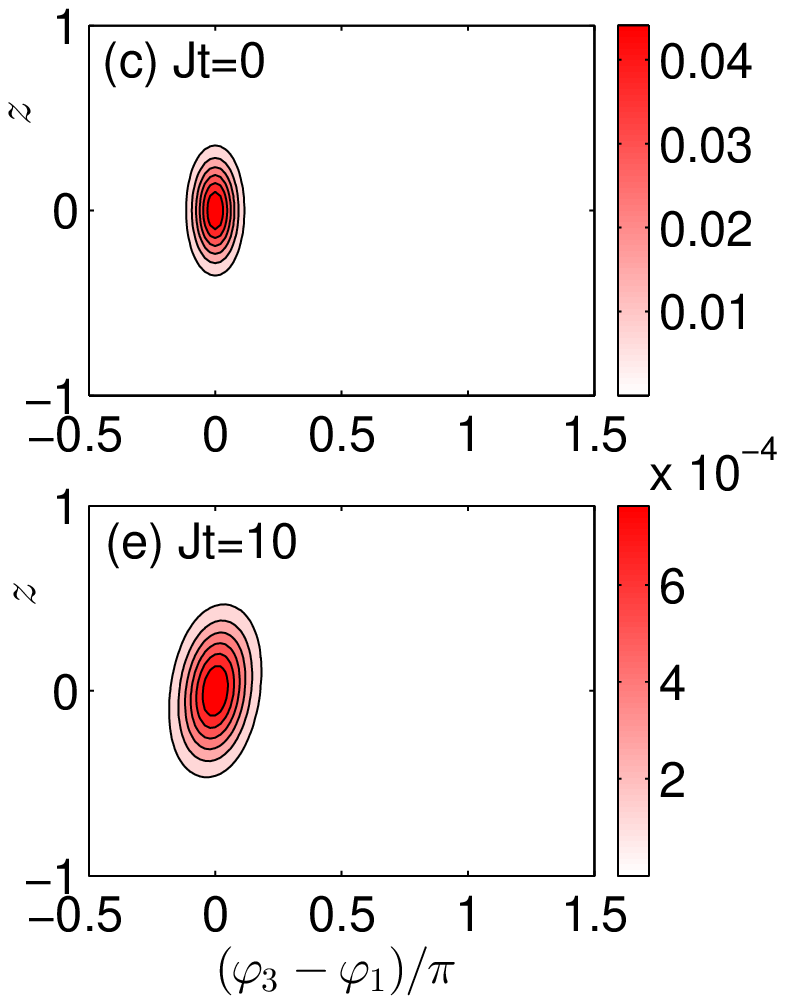}
\includegraphics[width=4cm, angle=0]{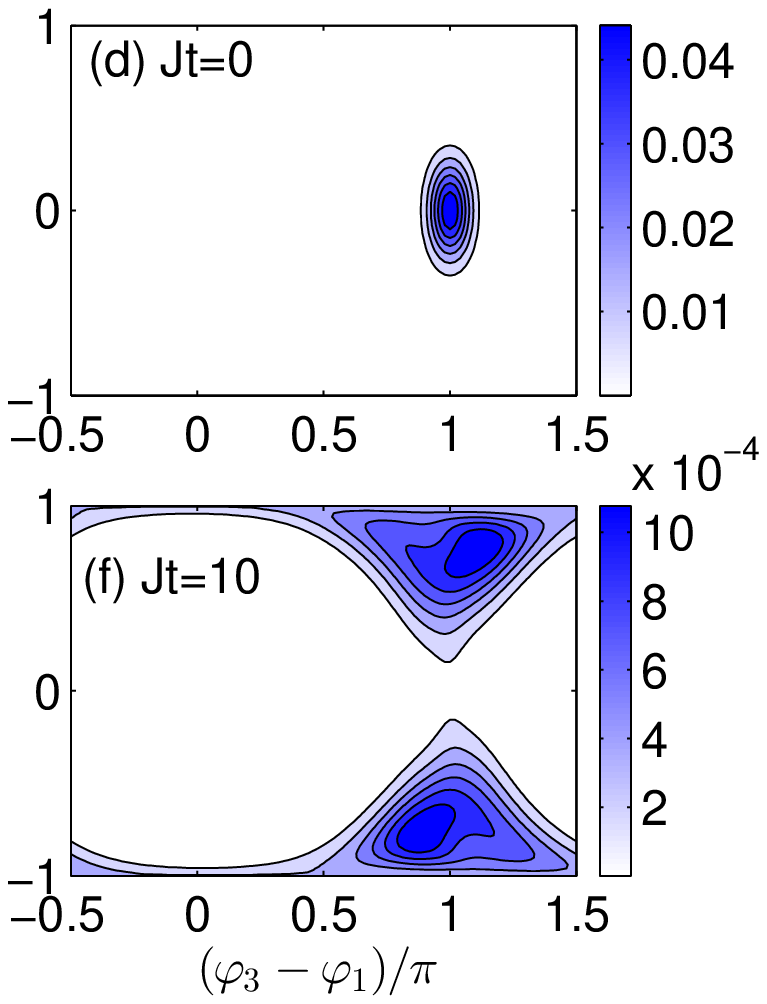}
\caption{\label{fig:husimi}
Semiclassical interpretation of breather state formation.
(a) Classical trajectories starting close to the symmetric state
$(\alpha_1,\alpha_2,\alpha_3) = (1,0,1)/\sqrt{2}$ (red) and
the anti-symmetric state
$(1,0,-1)/\sqrt{2}$ (blue).
(b) Meta-stable classical states as a function of the 
interaction strength. Plotted is the relative occupation 
at site 1. The icons on the right illustrate the population 
at the three sites and the dynamical stability of these states. 
Breathers emerge for $U N > 0.4 J$, $N$ being the total
atom number.
(c-f) The quantum dynamics of the $Q$ function follows
the classical phase space trajectories.
(c,e) A BEC with a symmetric wave function $\ket{\Psi_+}$ 
remains approximately pure.
(d,f) A BEC with an anti-symmetric wave function$\ket{\Psi_-}$ is 
coherently split into two parts forming the breather state.
The Husimi function $Q(\alpha_1,\alpha_2,\alpha_3)$
is plotted as a function of the population imbalance 
$z = (|\alpha_3|^2 - |\alpha_1|^2)/N$ and the 
relative phase  $\phi_3 - \phi_1$ for $\alpha_2 = 0$ and 
$|\alpha_1|^2 + |\alpha_3|^2 = N$.
Parameters as in Fig. \ref{fig:trimer}. 
}
\end{figure}

\section{Semiclassical interpretation}

The formation of breather states can be understood to a large extent
within a semi-classical phase space picture. 
Any quantum state can be represented by a quasi distribution function
on the associated classical phase space without loss of information
 \cite{Gard04}. 
In the following, we consider the Husimi function defined as 
$Q(\alpha_1,\alpha_2,\alpha_3;t) = \langle \alpha_1,\alpha_2,\alpha_3 | 
\hat \rho(t) | \alpha_1,\alpha_2,\alpha_3 \rangle$,
where $|\alpha_j\rangle$ is a Glauber coherent state in the $j$th well.
The dynamics of these 
distribution functions is to leading order given by a classical Liouville equation,
\be
   \frac{\partial Q}{\partial t} = - \sum_j \Big( \frac{\partial}{\partial \alpha_j} 
      \dot \alpha_j
       + \frac{\partial}{\partial \alpha_j^*} \dot \alpha_j^* \Big) Q
   + \mbox{noise}.
   \label{eqn:husimidyn}
\ee
Therefore the `classical' 
flow provides the skeleton of the quantum dynamics of the Husimi function,
whereas the quantum corrections vanish with increasing particle number as 
$1/N$ \cite{07phaseappl}. Figure~\ref{fig:husimi} (a) illustrates the 'classical' dynamics
which is given by the dissipative discrete Gross-Pitaevskii equation (DGPE) 
\be
   i \dot \alpha_j = -J (\alpha_{j+1} - \alpha_{j-1}) + U |\alpha_j|^2 \alpha_j
      - i \gamma_j \alpha_j/2 \, .
    \label{eqn:gp}
\ee
The figure shows the evolution of the population imbalance 
$z = (|\alpha_3|^2 - |\alpha_1|^2)/N$ and the relative phase  
$\Delta \varphi = \varphi_3 - \varphi_1$, where $\alpha_j = |\alpha_j| e^{i \varphi_j}$,
 for three different initial values. 
The trajectory with $\Delta \varphi=0$ (red) is dynamically stable,
such that it remains in the vicinity of the point $(z,\Delta \varphi) = (0,0)$ 
for all times.
In contrast, trajectories starting close to $(z,\Delta \varphi) = (0,\pi)$
converge to regions with either $z>0$ or $z <0$. These regions correspond
to self-trapped states, which are known from the non-dissipative 
case  \cite{Milb97,Smer97,Albi05}. 
For $\gamma_2 > 0$, these states become 
\emph{attractively stable}, which enables the dynamic formation of breather states.
Self-trapping occurs only if the interaction strength exceeds the critical value
$U_{\rm cr} = 0.4 \, J N^{-1}$ 
for the bifurcation shown in Fig.~\ref{fig:husimi} (b). The anti-symmetric
state $(z,\Delta \varphi) = (0,\pi)$ becomes unstable, whereas two attractively
stable self-trapping states emerge. The symmetric state remains marginally 
stable for all values of $U$. 

\begin{figure}[tb]
\centering
\includegraphics[width=5cm, angle=0]{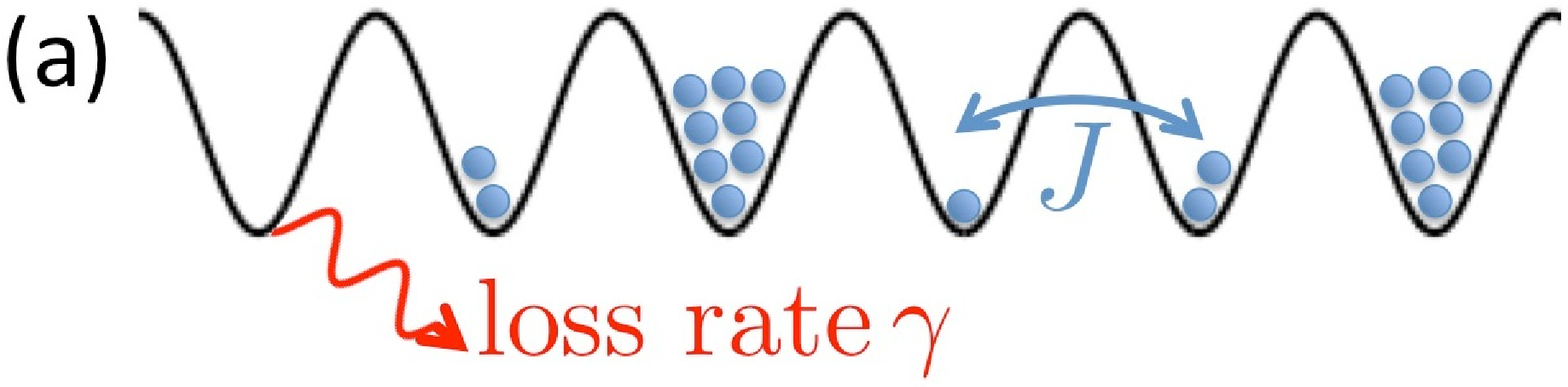}
\includegraphics[width=8cm, angle=0]{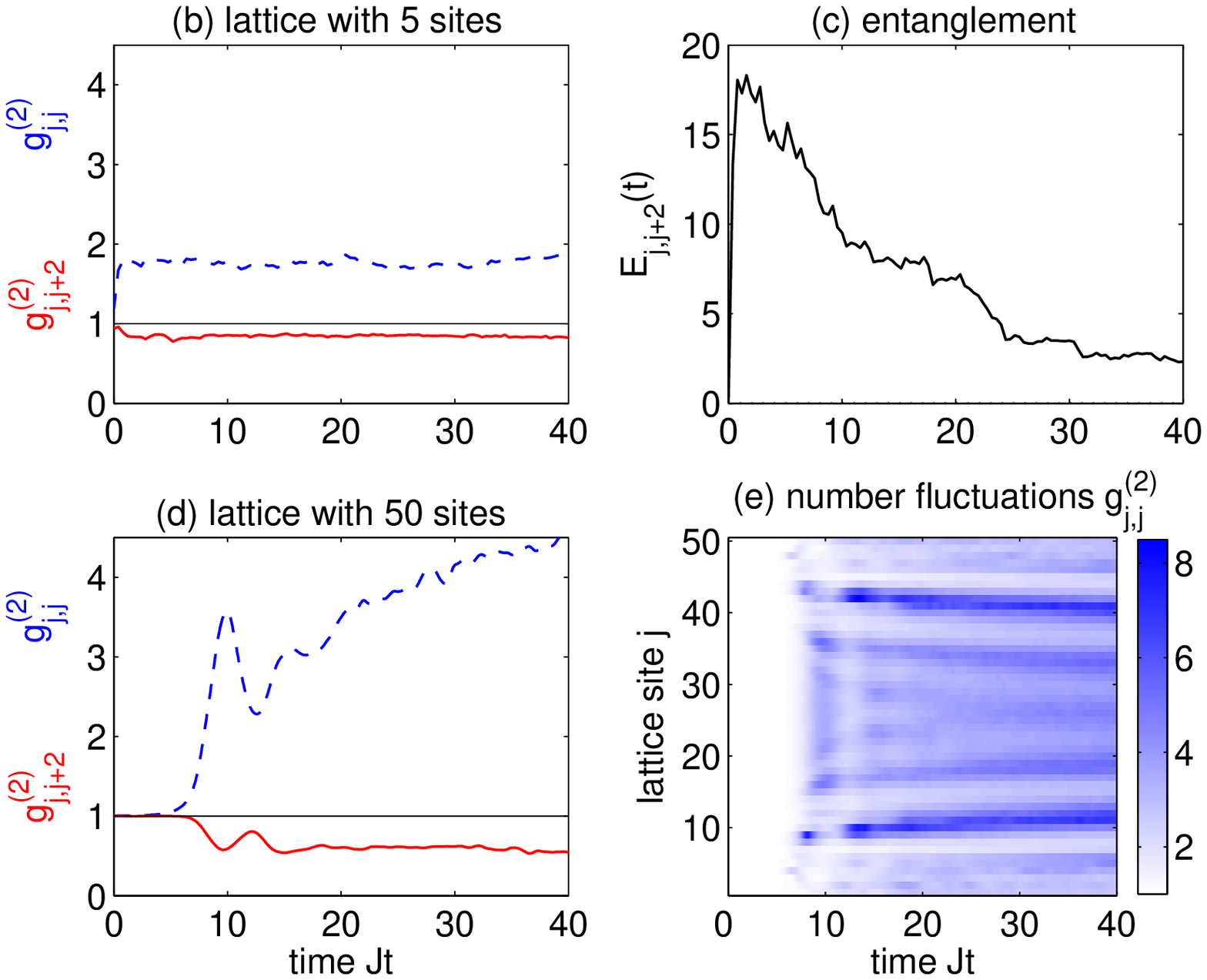}
\caption{\label{fig:lattice}
Generation of breather states in an optical lattice
by localized atom loss.
(a) Scheme of a possible experiment.
(b,d) The presence of large number fluctuations 
$g_{j,j}^{(2)}(t) > 1$ (dashed blue lines) and 
anti-correlations $g_{j,j+2}^{(2)}(t) < 1$ (solid red lines) 
indicates a breather state.
(c)The criterion $E_{j,j+2} > 0$ unambiguously 
proves the entanglement of the atoms
(cf.~Eq.~(\ref{eqn:ent_para})).
(e) The number fluctuations $g_{j,j}^{(2)}(t)$ increase
after a short transient period for all lattice sites $j$ in
a similar way.
Results are shown for (b,c) a small lattice with $M=5$ sites
and $j=3,$ $U = J$, $\gamma_1=0.2 J $, and $N(0)=16$ (solid lines)
and (d,e) an extended lattice with $M=50$ sites
and $j=25$, $UN(0)=25 J$, $\gamma_1=2 J $, and 
$N(0)/M=1000$ (dashed lines). 
The initial state is a pure BEC prepared at the band edge. 
}
\end{figure}

The corresponding quantum dynamics is shown in Fig.~\ref{fig:husimi} (c-f).
The Husimi functions of the symmetric initial state $\ket{\Psi_+}$
and the anti-symmetric initial state $\ket{\Psi_-}$ are localized around
$(z,\Delta \varphi) = (0,0)$ and $(z,\Delta \varphi) = (0,\pi)$, respectively,
as shown in Fig.~\ref{fig:husimi} (c,d).
As predicted by the DGPE, the symmetric state $\ket{\Psi_+}$
remains localized around $(z,\Delta \varphi) = (0,0)$ for all times.
On the contrary, the Husimi function of the anti-symmetric 
state $\ket{\Psi_-}$ flows to the self-trapping regions, such
that the final state is a superposition of two fragments with 
$z > 0$ and $z < 0$ -- a breather state (cf.~Fig.~\ref{fig:husimi} (e)).
The semi-classical picture predicts the fragmentation of the condensate
but, of course, cannot assert the coherence and thus the entanglement 
of the fragments which is a genuine quantum feature. However, it
correctly predicts the critical interaction strength for the emergence
of breather states.

\section{Extended lattices}

The entanglement protocol can be straightforwardly generalized to 
extended optical lattices, such that a realization is readily possible in 
ongoing experiments. First, a BEC is moved at constant speed \cite{sias07} or accelerated \cite{peik97}
to the edge of the first Brillouin zone such that the quantum state of
the atoms at $t=0$ is given by 
$|\Psi(0) \rangle  \sim (\sum \nolimits_j \psi_j \hat a_j^\dagger)^N \ket{0}$ 
with $\psi_j \sim (-1)^j$.
Then the atoms evolve freely according to the Master equation 
(\ref{eqn:master2}) subject to particle loss from lattice site $j=1$  
as illustrated in Fig.~\ref{fig:lattice} (a).
We simulate the dynamics for a small lattice with $M=5$ sites
and for an extended lattice with $M=50$ assuming periodic 
boundary conditions. In the latter case we use the truncated 
Wigner method \cite{Sina02}, which is especially suited for the 
 large filling factors considered here and describes
the deviation from a pure BEC state in contrast to a simple 
mean-field approach.

As a fingerprint for the dynamical generation of breather states we 
analyze the evolution of the number correlation functions as shown 
in Fig.~\ref{fig:lattice} (b,d).
After a short period of equilibration, the number fluctuations 
strongly increase, while the number correlations to the 
next-to-nearest neighboring site decrease. A similar picture 
emerges if we plot the $g^{(2)}_{j,j}$ functions for different 
lattice sites as in  Fig.~\ref{fig:lattice} (e).
As above, this fact shows that the atoms start to bunch in one or more 
breathers while the remaining lattice sites are essentially empty. 
In large lattices, breathers generally extend over more than one lattice site, 
such that anti-correlations $g^{(2)}_{j,k} < 1$ are observed only for the
next-to-nearest neighbors $|j-k| \ge 2$. The position of these breathers
is random due to quantum fluctuations. Hence, the final quantum 
state is a superposition of breathers at different sites, a macroscopically
entangled state. The rapid increase of the entanglement parameter
 (\ref{eqn:ent_para}) shown in Fig.~\ref{fig:lattice} (c) unambiguously
prooves the presence of many-particle entanglement for a small lattice 
with five sites. For long times $E_{3,5}(t)$ tends to zero again, simply
because all atoms have decayed from the lattice.
Breathers appear only when the interaction strength $Un$ exceeds a 
critical value, which can be infered from semiclassical arguments.
This transition can be interpreted
as a dynamical phase transition \cite{Ng09}
and will be analyzed in detail in a forthcoming article \cite{12dime2}.

\section{Conclusion}

Engineering dissipation is a promising new direction in the control
of complex quantum systems. We have shown that 
an elementary dissipation process, the localized loss of particles,
together with repulsive interactions, is sufficient to create 
macroscopically entangled states of ultracold atoms in optical
lattices. 
The quantum state is a coherent superposition of atoms
bunched at different lattice sites.
We have discussed the properties of these `breather states' in detail,
 including
entanglement, decoherence and applications in precision 
quantum interferometry.  
A semiclassical interpretation of breather state formation has 
revealed the connection to a classical bifurcation of the associated
mean-field dynamics.

Breather states are significantly different from squeezed entangled states, where  interactions reduce (`squeeze') the number variance in a well of the lattice \cite{Gros10,Sore01}. 
In a breather state, a well is either occupied by a large number of atoms or empty, giving rise to a large number variance.
The entanglement enables precision  metrology beyond the standard quantum limit using protocols introduced for optical NOON states.

The introduced protocol can be readily implemented 
experimentally with ultracold atoms in optical lattices. 
Localized access to the lattice can be realized either
optically or by a focussed electron beam \cite{Albi05,Gros10,Bakr09,Sher10,Geri08,Wurt09}.
Macroscopically entangled breather states are then formed dynamically as meta-stable states of the dissipative quantum dynamics. 
This protocol is very favorable as no fine-tuning of parameters 
is needed and the entanglement persists as long as enough
atoms remain in the lattice.
As particle loss is an elementary dissipation process,
the effects discussed here may be  important
for a variety of different physical systems, as for instance, 
optical fiber experiments \cite{Rege11}.

We thank O.~G\"uhne for stimulating discussions.
We acknowledge financial support by the Max Planck Society, the Deutsche Forschungsgemeinschaft 
via FOR760, the individual grant WI 3426/3-1 and the HGSFP (GSC 129/1).

\section{Appendix}

In this appendix we present a detailed derivation of the
entanglement criterion (\ref{eqn:ent_para}). 
This result generalizes established entanglement
criteria in terms of spin squeezing~\cite{Sore01} and is derived
in a similar way. In contrast to spin squeezing inequalities, it shows that
a state is entangled if the varianace (\ref{eqn:ent_para}), defined in the text, is
\emph{larger} than a certain threshold value.
We assume that the many-body quantum state $\hat \rho$ is 
decomposed into a mixture of pure states
\be
   \hat \rho = \sum \nolimits_a p_a \hat \rho_a
    =  \sum \nolimits_a p_a \ket{\psi_a}\bra{\psi_a},
   \label{eqn:rhodecompose}
\ee
where every pure state $\hat \rho_a = \ket{\psi_a}\bra{\psi_a}$ 
has a fixed particle number $N_a$. Now we proof that the 
entanglement parameter (\ref{eqn:ent_para}) is negative, $E_{j,k}<0$, 
for every separable state such that a value $E_{j,k}>0$ 
unambiguously reveals the presence of
many-particle entanglement.

We start with pure states $\hat \rho_a$.
If such a pure state $\hat \rho_a$ is separable, it can be written 
as a tensor product of single particle states
\be 
   \hat \rho_a =   \hat \rho_a^{(1)} \otimes   \hat \rho_a^{(2)} 
           \otimes \cdots \otimes   \hat \rho_a^{(N_a)}.
   \label{eqn:psstate}
\ee
We furthermore introduce the abbreviation
$\hat S_\pm := \hat n_j \pm \hat n_k$.
This operator is also written as a tensor product of single-particle
operators
\bea
   \hat S_\pm = \sum_{r=1}^N \eye \otimes \cdots \otimes \eye 
       \otimes \hat s_\pm^{(r)} \otimes \eye \otimes \cdots \otimes \eye,
\eea
where the superscript $(r)$ denotes that the single-particle operator $\hat s_\pm^{(r)}$ 
acts on the $r$th atom. The single-particle operators are given by
$\hat s_\pm = \ket{j}\bra{j} \pm  \ket{k}\bra{k}$,
where $\ket{j}$ is the quantum state where the particle is 
localized in site $j$.

For a separable pure state  $\hat \rho_a$, the expectation 
values of the population imbalance 
$ \langle \hat S_- \rangle_a = \tr [\hat \rho_a \hat S_-]$
and its square can be expressed as
\bea
    \langle \hat S_- \rangle \!\!\! &=& \!\!\!
      \sum_{r=1}^N \tr\left[ \rho^{(r)} \hat s_-^{(r)}  \right], \\
  \langle \hat S_-^2 \rangle \!\!\! &=& \!\!\!
     \sum_{r \neq q}^N \tr \left[ (\rho^{(r)} \otimes \rho^{(q)} )
            (\hat s_-^{(r)} \otimes \hat s_-^{(q)}) \right] 
  \\  && \qquad
         + \sum_{r=1}^N  \tr\left[ \rho^{(r)} \hat s_-^{(r)2}  \right] \nn \\
   &=& \!\!\! \sum_{r,q=1}^N \tr\left[ \rho^{(r)} \hat s_-^{(r)}  \right] 
              \tr\left[ \rho^{(q)} \hat s_-^{(q)}  \right] \nn \\
      && \!\!\!  - \sum_{r=1}^N \tr\left[ \rho^{(r)} \hat s_-^{(r)}  \right] 
              \tr\left[ \rho^{(r)} \hat s_-^{(r)}  \right] 
           + \sum_{r=1}^N \tr\left[ \rho^{(r)} \hat s_-^{(r)2}  \right] \nn \\
      &=& \!\!\! \langle \hat S_- \rangle^2 +
           \sum_{r=1}^N  \tr\left[ \rho^{(r)} \hat s_-^{(r)2}  \right]   
              - \left\{ \tr\left[ \rho^{(r)} \hat s_-^{(r)}  \right] \right\}^2. \nn
\eea
Using $\tr[ \rho^{(r)} \hat s_-^{(r)2} ] = \tr[ \rho^{(r)} \hat s_+^{(r)}]$ 
we thus find that every pure products state $\hat \rho_a$ satisfies 
the condition
\be
   \langle \hat S_-^2 \rangle_a - \langle \hat S_- \rangle^2_a \le \langle \hat S_+ \rangle_a \, .
\ee

If the total quantum state $\hat \rho$ is separable, such that 
it can be written as a mixture of separable pure states 
(\ref{eqn:rhodecompose}), the expectation values are given by
\bea
   && \langle \hat S_-^2 \rangle = \sum_a p_a \langle \hat S_-^2 \rangle_a  
    \le \langle \hat S_+ \rangle + \sum_a p_a  \langle \hat S_- \rangle_a^2 \nn \\
   && \langle \hat S_- \rangle^2 = \sum_{a,b} p_a p_b \langle \hat S_- \rangle_a \langle \hat S_- \rangle_b \\
   && \qquad = \sum_a p_a  \langle \hat S_- \rangle_a^2
            - \frac{1}{2} \sum_{a,b} p_a p_b \left[ \langle \hat S_- \rangle_a - \langle \hat S_- \rangle_b \right]^2. \nn
\eea
We thus find that every separable quantum state satisfies the following inequality
for the variance  of the population imbalance $\hat S_-$: 
\be
   \langle \hat S_-^2 \rangle - \langle \hat S_- \rangle^2
   \le \langle \hat S_+ \rangle +
    \frac{1}{2} \sum_{a,b} p_a p_b \left[ \langle \hat S_- \rangle_a - \langle \hat S_- \rangle_b \right]^2.
\ee
This inequality for separable states can be rewritten as
$E_{j,k} < 0$
in terms of the entanglement parameter (\ref{eqn:ent_para}).


\end{document}